\documentclass[aps,twocolumn,showpacs,superscriptaddress]{revtex4-1}
\usepackage[english]{babel}
\usepackage{graphicx,dcolumn,bm,amssymb,amsmath,amsfonts,amsthm,latexsym}
\usepackage{float,subfig,slashed,hyperref}
\hypersetup{
  linktocpage  = true,
  colorlinks   = true,
  urlcolor     = red,
  linkcolor    = black,
  citecolor    = blue
}
\begin{document}
\widetext

\title{Meson properties and phase diagrams\\ in a SU(3) nlPNJL model with lQCD-inspired form factors}
\author{J.~P.~Carlomagno}
\email{carlomagno@fisica.unlp.edu.ar}
\affiliation{IFLP, CONICET $-$ Dpto.\ de F\'{\i}sica, Universidad Nacional de La Plata, C.C. 67, 1900 La Plata, Argentina}
\affiliation{CONICET, Rivadavia 1917, 1033 Buenos Aires, Argentina}

\begin{abstract}
We study the features of a nonlocal SU(3) Polyakov-Nambu-Jona-Lasinio model that includes wave function renormalization.
Model parameters are determined from vacuum phenomenology considering lattice QCD-inspired nonlocal form factors.
Within this framework we analyze the properties of light scalar and pseudoscalar mesons at finite temperature and chemical potential determining characteristics of deconfinement and chiral restoration transitions.
\end{abstract}

\maketitle

\section{Introduction}
\label{intro}

The strong interaction among quarks depends on their color charge. 
When quarks are placed in a medium, this charge is screened due to density and temperature effects~\cite{Fukushima:2010bq}. 
If either of these increase beyond a certain critical value, the interactions between quarks no longer confine them inside hadrons.
This is usually referred to as the deconfinement phase transition.
In addition, another transition takes place when the realization of chiral symmetry shifts from a Nambu-Goldstone to a Wigner-Weyl phase.
Based on lattice QCD (lQCD) evidence at zero chemical potential~\cite{Bazavov:2016uvm} one expects these two phase transitions to occur at approximately the same temperature. 
At finite density, in principle, they could arise at different critical temperatures, leading to a quarkyonic phase, in which the chiral symmetry is restored while quarks and gluons remain confined.

Although QCD is a first principle theory of hadron interactions, it has the drawback of being a theory where the low energy regime is not available using standard perturbative methods.
This problem can be addressed from first principles through lattice calculations~\cite{Allton:2003vx, Allton:2005gk, Fodor:2004nz, Aoki:2005vt, Karsch:2003jg}. 
However, this approach has difficulties when dealing with small current quark masses and/or finite chemical potential. 
Thus, some of the present knowledge about the behavior of strongly interacting matter arises from the study of effective models, which offer the possibility to get predictions of the transition features at regions that are not accessible through lattice techniques.

Here we will concentrate on one particular class of effective theories, viz. the nonlocal Polyakov$-$Nambu$-$Jona-Lasinio (nlPNJL) models (see~\cite{Carlomagno:2013ona} and references therein), in which quarks interact through covariant nonlocal chirally symmetric four and six point couplings in a background color field and gluon self-interactions are effectively introduced by a Polyakov loop effective potential.
These approaches, which can be considered as an improvement over the (local) PNJL model, offer a common framework to study both the chiral restoration and deconfinement transitions. 
In fact, the nonlocal character of the interactions leads to a momentum dependence in
the quark propagator that can be made consistent~\cite{Noguera:2008} with
lattice results.

Some previous works have addressed the study of meson properties and/or phase transitions using nlPNJL models with Gaussian nonlocal form factors, for specific Polyakov potentials~\cite{Contrera:2009hk}.
These functional forms can be improved, since it is possible to choose model parameters and momentum dependences for the form factors so as to fit the quark propagators obtained in lattice QCD.
The aim of this work is to extend the above references to finite chemical potential with lQCD-inspired form factors, and determine several properties of light mesons (masses, mixing angles, decay constants) at zero and finite temperature, analyzing the compatibility with the corresponding phenomenological values. 
In addition, we will study the deconfinement and chiral restoration phase transitions at finite temperature and density, obtaining the critical temperatures and chemical potentials, and sketching the corresponding phase diagram.

This article is organized as follows. In Sect.~\ref{nlpnjl} we present the general formalism for a finite temperature and density system. 
The numerical and phenomenological analyses at zero and finite temperature for several meson properties are included in Sect.~\ref{thermo}. 
In Sect.~\ref{pd} we present the phase diagrams for different Polyakov loop potentials and discuss the phase transition features.
Finally, in Sect.~\ref{summary} we summarize our results and present the conclusions. 

\section{Thermodynamics}
\label{nlpnjl}

Let us consider an Euclidean action for a three flavor quark model with nonlocal four and six point couplings~\cite{Carlomagno:2013ona},
\begin{eqnarray}
\label{se}
S_E &=& \int d^4x \ \left\{ \overline{\psi}(x)(-\imath \slashed D + \hat m)\psi(x) \right. \nonumber \\ 
	&&\left. -\frac{G}{2}\left[j_a^S(x)j_a^S(x)+j_a^P(x)j_a^P(x)+j^r(x)j^r(x)\right] \right. \nonumber\\
    &&\left. -\frac{H}{4} A_{abc}\left[ j_a^S(x)j_b^S(x)j_c^S(x)-3j_a^S(x)j_b^P(x)j_c^P(x)\right] \right. \nonumber \\
    && \left. + \ {\cal U} \,[{\mathcal A}(x)] \right\} \ .
\end{eqnarray}
Here, $\psi(x)$ is the $N_f=3$ fermion triplet $\psi = (u\ d\ s)^T$, and $\hat m={\rm diag}(m_u,m_d,m_s)$ is the current quark mass matrix. We will work in the isospin symmetry limit, assuming $m_u=m_d$. 
The fermion currents are given by
\begin{eqnarray}
\label{currents}
j_a^s(x) &=& \int d^4z\; g(z)\,
\overline{\psi}\left(x+\frac{z}{2}\right)\lambda_a
\psi\left(x-\frac{z}{2}\right)
\ , \nonumber\\
j_a^p(x) &=& \int d^4z\; g(z)\,
\overline{\psi}\left(x+\frac{z}{2}\right)\imath \lambda_a \gamma_5
\psi\left(x-\frac{z}{2}\right)
\ , \nonumber\\
j^r(x)   &=& \int d^4z\; f(z)\, \overline{\psi}\left(x+\frac{z}{2}\right)
\frac{\imath \overleftrightarrow{\slashed \partial}}{2\kappa}
\psi\left(x-\frac{z}{2}\right)\ ,
\end{eqnarray}
where $f(z)$ and $g(z)$ are covariant form factors responsible for the nonlocal character of the interactions, and $\lambda_a$, $a=0,...,8$, are the standard Gell-Mann matrices, plus $\lambda_0=\sqrt{2/3}\,\mathbf{1}_{3\times 3}$. 
The relative weight of the interaction driven by $j^r(x)$, responsible for the quark wave function renormalization (WFR), is controlled by the parameter $\kappa$. 

The interaction between fermions and color gauge fields $G_\mu^a$ takes place through the covariant derivative in the fermion kinetic term,
$D_\mu\equiv
\partial_\mu - \imath {\mathcal A}_\mu$, where ${\mathcal A}_\mu = g\,
G_\mu^a \lambda^a/2$.
In this effective model we will assume that fermions move on a static and constant background gauge field $\phi$. 
The traced Polyakov loop (PL) $\Phi$, which in the infinite quark mass limit can be taken as the order parameter for confinement~\cite{tHooft:1977nqb, Polyakov:1978vu}, is given by
\begin{equation}
\Phi=\frac{1}{N_c}{\rm Tr}\ {\cal P}\,\exp\left(i\int_0^{1/T}\!dx_4\,\phi\right).
\end{equation}

The effective gauge field self-interactions in Eq.~(\ref{se}) are given by the Polyakov-loop potential ${\cal U}\,[A(x)]$. 
At finite temperature $T$, it is usual to take for this potential a functional form based on properties of pure gauge QCD. 
The potential is constrained by the condition of reaching the Stefan-Boltzmann limit at $T \rightarrow \infty$ and by requiring the presence of a first-order phase transition at a given temperature $T_0$~\cite{Schaefer:2007pw}.
In the presence of dynamical flavors this parameter has to be rescaled from the pure gauge transition temperature (of about $270$~MeV) toward values around $200$~MeV~\cite{Carlomagno:2013ona}.
In addition, it has been argued that $T_0$ should change with the chemical potential $\mu$ as~\cite{Schaefer:2007pw,Ciminale:2007sr}
\begin{equation}
T_0 = T_{\tau}\ e^{-1/\alpha_0 b(\mu)} \ ,
\label{T0ofmu}
\end{equation}
where $T_{\tau} = 1.77$~GeV, $\alpha_0 = 0.304$ and $b(\mu) = 1.508 - 32/\pi\ (\mu/T_{\tau})^2$. This dependence is motivated by the calculation of hard dense loop and hard thermal loop contributions to the effective charge~\cite{Lebellac:1996}.

A possible Ansatz for the PL potential is given by a logarithmic form based on the Haar measure of the SU(3) color group, namely~\cite{Roessner:2006xn}
\begin{eqnarray}
\frac{{\cal{U}}_{\rm log}(\Phi ,T)}{T^4}  &&\ = 
\ -\,\frac{1}{2}\, a(T)\,\Phi^2 \;+ \nonumber \\
&&\;b(T)\, \log\left(1 - 6\, \Phi^2 + 8\, \Phi^3
- 3\, \Phi^4 \right) \ ,
\label{ulog}
\end{eqnarray}
where
\begin{eqnarray}
&& a(T) = a_0 +a_1 \left(\dfrac{T_0}{T}\right) + a_2\left(\dfrac{T_0}{T}\right)^2 \ , \nonumber \\
&& b(T) = b_3\left(\dfrac{T_0}{T}\right)^3 \ . \nonumber
\label{log}
\end{eqnarray}

Another widely used potential is that given by a polynomial function based on a Ginzburg-Landau ansatz~\cite{Ratti:2005jh,Scavenius:2002ru}:
\begin{eqnarray}
\frac{{\cal{U}}_{\rm poly}(\Phi ,T)}{T ^4} \ = \ -\,\frac{b_2(T)}{2}\, \Phi^2
-\,\frac{b_3}{3}\, \Phi^3 +\,\frac{b_4}{4}\, \Phi^4 \ ,
\label{upoly}
\end{eqnarray}
where
\begin{eqnarray}
b_2(T) = a_0 +a_1 \left(\dfrac{T_0}{T}\right) + a_2\left(\dfrac{T_0}{T}\right)^2
+ a_3\left(\dfrac{T_0}{T}\right)^3\ . \nonumber
\label{pol}
\end{eqnarray}
Numerical values for parameters $a_i$ and $b_i$ in these potentials can be found in Refs.~\cite{Roessner:2006xn,Ratti:2005jh,Scavenius:2002ru}.

\subsection*{Mean Field Approximation}

To determine the QCD phase diagram in the $T-\mu$ plane we consider the thermodynamic potential per unit volume at mean field level (MF).
We proceed by using the standard Matsubara formalism. 
Following the same procedure as in Refs.~\cite{Carlomagno:2013ona,Scarpettini:2003fj,GomezDumm:2001fz}, we perform a standard bosonization of the fermionic theory, Eq.~(\ref{se}), introducing scalar fields $\sigma_a(x)$, $\zeta(x)$ and pseudoscalar fields $\pi_a(x)$, with $a=0,...,8$. 
We obtain
\begin{equation}
\Omega^{\rm MF} (T,\mu)  \ = \ \Omega^{\rm reg} + \Omega^{\rm free} +
\mathcal{U}(\Phi,T) + \Omega_0 \ , \nonumber
\end{equation}
where
\begin{widetext}
\begin{eqnarray}
\Omega^{\rm reg} &=& -2\, \sum_{c,f}\ T \sum_{n=-\infty}^{\infty} \int \dfrac{d^3p}{(2\pi)^3} \log
\left[\dfrac{p_{nc}^2 + M_f^2(p_{nc})}{Z^2(p_{nc})\, (p_{nc}^2 + m^2_f)} \right]
 - \left(\bar\zeta\, \bar R + \dfrac{G}{2}\,\bar R^2 +
\dfrac{H}{4}\,\bar S_u\, \bar S_d \, \bar S_s \right)
- \dfrac{1}{2}\, \sum_f \left( \bar \sigma_f \bar S_f
+ \dfrac{G}{2}\, \bar S_f^2 \right)\ ,
\nonumber \\
\Omega^{\rm free} &=& -2\, T \sum_{c,f} \sum_{s=\pm 1} \int
\dfrac{d^3p}{(2\pi)^3}\; {\rm Re}\,\log
\left[1+\exp\left(-\frac{\epsilon_{fp} + s\,(\mu\,+\imath\, \phi_c)}{T}
\right)\right]\ .
\end{eqnarray}
\end{widetext}
Here we have defined $p_{nc}^2 = [(2n+1)\pi T+\phi_c-\imath\ \mu]^2 + \vec p^{\;2}$, $\epsilon_{fp}=\sqrt{\vec p^{\;2} + m_f^2}$. 
The sums over color and flavor indices run over $c=r,g,b$ and $f=u,d,s$, respectively, and the color background fields are $\phi_r = - \phi_g = \phi$, $\phi_b = 0$. 
The term $\Omega^{\rm free}$ is the regularized expression for the thermodynamical potential of a free fermion gas, while $\Omega_0$ is just a constant that fixes the value of the thermodynamical potential at $T=\mu=0$. 

The functions $M_f(p)$ and $Z(p)$ correspond to momentum-dependent effective masses and WFR of the quark propagators. 
In terms of the model parameters and form factors, these are given by
\begin{eqnarray}
M_f(p) &=& Z(p)\, \left[ m_f\, +\, \bar \sigma_f\, g(p)\right] \ ,
\nonumber\\
Z(p) &=& \left[ 1\,-\,\frac{\bar \zeta}{\kappa}\,f(p)\right]^{-1}\ ,
\label{mz}
\end{eqnarray}
where $\bar \sigma_f$ and $\bar\zeta$, are the vacuum expectation values of the scalar fields introduced to bosonize the fermionic theory. 
We use the stationary phase approximation, where the path integrals over the corresponding auxiliary fields $S_f$ and $R$ are replaced by the arguments evaluated at the minimizing values $\bar S_f$ and $\bar R$. 
The procedure is similar to that carried out in Ref.~\cite{Scarpettini:2003fj}, where more details can be found.
From the minimization of this regularized thermodynamic potential it is possible to obtain a set of coupled gap equations that determine $\bar \sigma_f$, $\bar\zeta$ and $\phi$ at a given temperature $T$ and chemical potential $\mu$
\begin{equation}
\dfrac{\partial \Omega^{\rm MF} (T,\mu)}{\partial \bar \sigma_f} = 
\dfrac{\partial \Omega^{\rm MF} (T,\mu)}{\partial \bar \zeta} = 
\dfrac{\partial \Omega^{\rm MF} (T,\mu)}{\partial \phi } = 0 \ . \nonumber
\label{gap_eq}
\end{equation}

To characterize the chiral and deconfinement phase transitions it is necessary to define the corresponding order parameters. 
It is well known that the chiral quark condensates $\langle \bar q q\rangle$ are appropriate order parameters for the restoration of the chiral symmetry. Their expression can be obtained by varying the MF partition function with respect to the  current quark masses. 
In general, these quantities are divergent, and can be regularized by subtracting the free quark contributions. 
Therefore, it is usual to define a subtracted chiral condensate, normalized to its value at $T=0$, as
\begin{equation}
\langle \bar q q\rangle_{\rm sub} \ = \ \dfrac{\langle \bar u u \rangle \,
-\, \frac{m_u}{m_s} \, \langle \bar s s\rangle}{\langle \bar u u \rangle_{0} \,
- \, \frac{m_u}{m_s} \, \langle \bar s s \rangle_{0}} \ .
\label{qq_sub}
\end{equation}

Regarding the description of the deconfinement transition, a crucial role is played by the center symmetry $Z(N)$ of the pure Yang-Mills theory. 
As stated, we will take as the corresponding order parameter the trace of the Polyakov line, given by~\cite{Diakonov:2004kc}
\begin{equation}
\Phi = \dfrac{1}{3}\ [1 + 2 \cos(\phi/T)] \ .
\label{Pkv_line}
\end{equation}
If $\Phi =0$, $Z(N)$ symmetry is manifest, and this situation indicates confinement. 
Above the critical temperature one has $\Phi \neq 0$, therefore, the symmetry is broken, which corresponds to the deconfined phase.
For the light quark sector, $\Phi$ turns out to be an approximate order parameter for the deconfinement transition in the same way that the chiral quark condensate is an approximate order parameter for the chiral symmetry restoration outside the chiral limit.

\subsection*{Observables beyond mean field}

The study of meson properties at finite temperature has to be carried out beyond mean field.
The quadratic contribution (in powers of mesonic fluctuations) to the thermodynamical potential is given by
\begin{eqnarray}
\label{spiketa}
\Omega^{\rm quad} &=& \dfrac{1}{2}\ T \sum_{k} \int \dfrac{d^3q}{(2\pi)^3} 
G_M(q_k^2)\  \phi_M(q_k)\, \phi_M(-q_k) \ , \nonumber
\end{eqnarray}
where $\phi_M$ correspond to the meson fields in the SU(3) charge basis. Here $M$ labels the scalar and pseudoscalar mesons in the lowest mass realization, plus the $\zeta$ field, and $q_k=(\vec{q},\nu_k)$, where $\nu_k=2k \pi T$, are bosonic Matsubara frequencies. 

Meson masses are then given by the equations~\cite{Carlomagno:2013ona,Contrera:2009hk}
\begin{equation}
\label{Ges}
G_M(-m_M^2)\ =\ 0 \ .
\end{equation}
The mass values determined by these equations at $\vec q=(0,0,\imath\ m_M)$ with $k=0$ correspond to the spatial masses of the mesons’ zeroth Matsubara modes, their inverses describing the persistence lengths of these modes at equilibrium with the heat bath.

The one-loop functions $G_M$ can be written in terms of the coupling constants $G$ and $H$, the mean field values $\bar S_{u,s}$ and quark loop integrals that prove to be ultraviolet convergent owing to the asymptotic behavior of the nonlocal form factors. 

For the pseudoscalar meson sector one can also evaluate mixing angles and weak decay constants.
The latter are given by the matrix elements of the axial currents between the vacuum and the physical meson states.
Since the $I=0$ states get mixed, it is necessary to introduce mixing angles $\theta_\eta$ and $\theta_{\eta'}$ to diagonalize this coupled sector.

Calculation details, together with the definitions of above quantities at zero temperature, can be found in Ref.~\cite{Carlomagno:2013ona}.
Our aim is to extend here those results to a finite temperature system.

\section{Meson phenomenology}
\label{thermo}

\subsection*{Model parameters and form factors} 

The model includes five free parameters, namely the current quark masses $m_{u,s}$ and the coupling constants $G$, $H$ and $\kappa$. 
In addition, one has to specify the form factors $f(z)$ and $g(z)$ in the nonlocal fermion currents of  Eq.~(\ref{currents}). 
Here, we will consider for the form factors a momentum dependence based on lQCD results for the quark propagators. 
Therefore, following the analysis of Ref.~\cite{Bowman:2002bm}, we parametrize the effective mass $M_f(p)$ as
\begin{equation}
M_f(p) \ = \ m_f \, + \, \alpha_m\, f_m(p)\ ,
\label{fm}
\end{equation}
where
\begin{equation}
f_m(p) \ = \ \frac{1}{1 + (p^2/\Lambda_0^2)^\alpha}\ ,
\label{fm2}
\end{equation}
with $\alpha = 3/2$.
On the other hand, for the WFR we use the parametrization~\cite{Noguera:2005ej,Noguera:2008}
\begin{equation}
Z (p) \ = \ 1 \, - \, \alpha_z\, f_z(p)\ ,
\label{fz}
\end{equation}
where
\begin{equation}
f_z(p) \ = \ \frac{1}{\left(1 + p^2/\Lambda_1^2\right)^{\beta}}\ .
\end{equation}
It is found that lQCD results favor a relatively low value for the exponent, hence we take here $\beta = 5/2$, which is the smallest exponent compatible with the ultraviolet convergence of the loop integrals.
The coefficients $\alpha_m$ and $\alpha_z$ can be expressed in terms of the mean field values $\bar{\sigma}_u$ and $\bar{\zeta}$ (see Eq.~(\ref{mz})).
From Eqs.~(\ref{mz}), Eq.~(\ref{fm}) and Eq~(\ref{fz}) one can relate the functions $f(p)$ and $g(p)$ to $f_m(p)$ and $f_z(p)$. 

\begin{figure*}[t]
\centering
\subfloat{\includegraphics[width=0.45\textwidth]{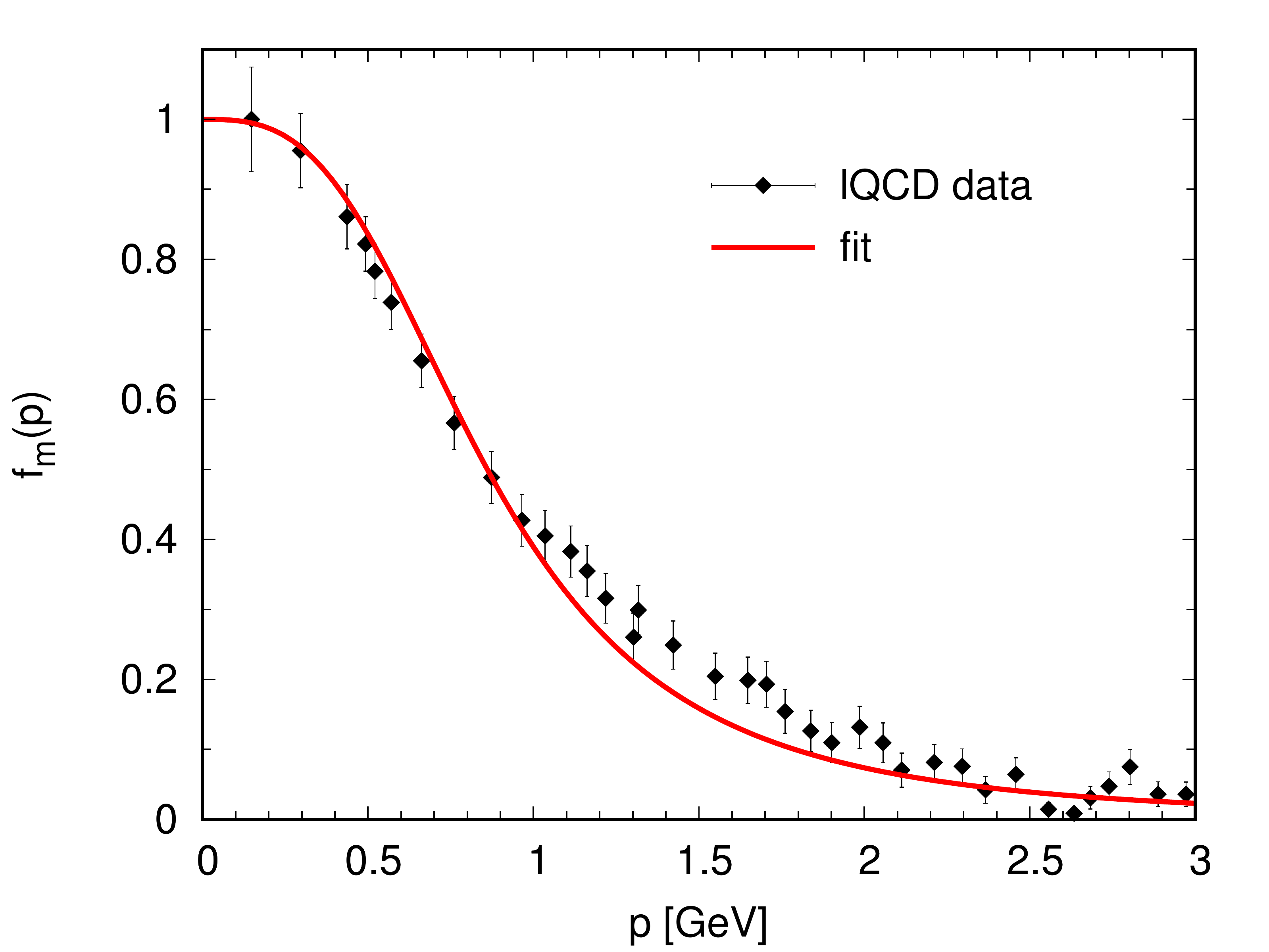}}
\subfloat{\includegraphics[width=0.45\textwidth]{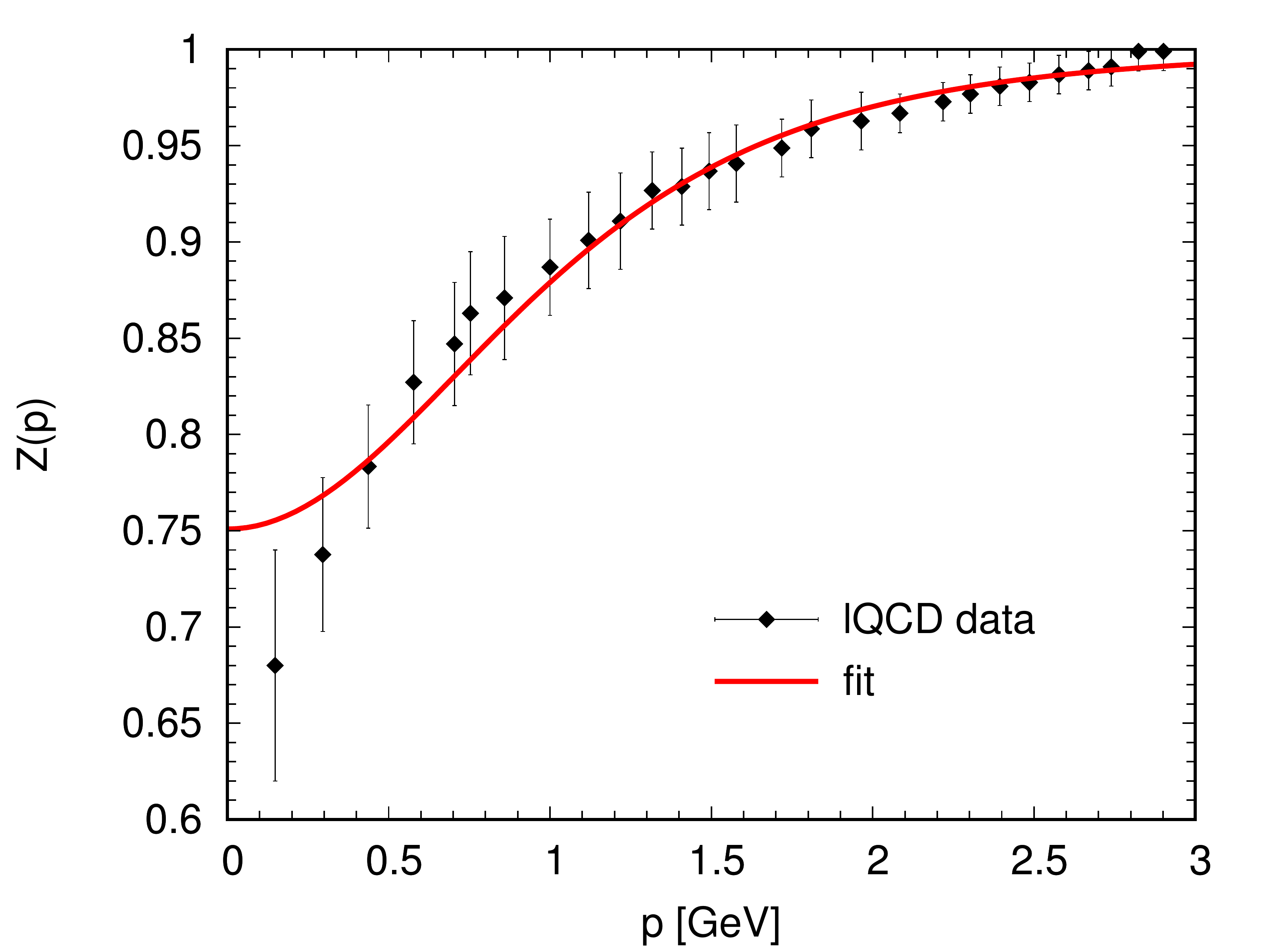}}
\caption{\small{Fit to lattice data from Ref.~\cite{Parappilly:2005ei} for the functions $f_m(p)$ and $Z(p)$, Eqs.~(\ref{fm2}) and~(\ref{fz}).}}
\label{fig:ff}
\end{figure*}

Given the form factor functions, it is possible to set the model parameters to reproduce the observed meson phenomenology. 
To the above mentioned free parameters ($m_u$, $m_s$, $G$, $H$ and $\kappa$) one has to add the cutoffs $\Lambda_0$ and $\Lambda_1$, introduced in the form factors.  
Through a fit to lQCD results quoted in Ref.~\cite{Parappilly:2005ei} for the functions $f_m(p)$ and $Z(p)$, we obtain
\begin{eqnarray*}
&&\Lambda_0 = 861 {\rm \ MeV}\ , \quad
\Lambda_1 = 1728 {\rm \ MeV}\ , \quad
\alpha_z = -0.2492\ .
\label{fit}
\end{eqnarray*}
The curves corresponding to the functions $f_m(p)$ and $Z(p)$, together with $N_f=2+1$ lattice data are shown in Fig.~\ref{fig:ff}.
The fit has been carried out considering results up to 3~GeV.

The remaining five parameters can be determined by requiring that the model reproduces the empirical values of four physical quantities and the value of $\alpha_z$ obtained from the fit. 
We have taken as inputs the masses of the pseudoscalar mesons $\pi$, $K$ and $\eta^{\prime}$, and the pion weak decay constant $f_{\pi}$. 
The obtained values of the model parameters are quoted in Table~\ref{tab:param}.
\begin{table}[h]
\begin{center}
\begin{tabular*}{0.2\textwidth}{@{\extracolsep{\fill}} c c }
\hline
\hline
Parameter &  Value \\
\hline
$m_u$ [MeV] & 2.38  \\
$m_s$ [MeV] & 61.45  \\
$G \Lambda_0^2$  & 14.03 \\
$H \Lambda_0^5$ & -158.70  \\
$\kappa$ [GeV] & 12.45  \\
\hline
\hline
\end{tabular*}
\caption{\small{Model parameter values.}}
\label{tab:param}
\end{center}
\end{table}

\subsection*{Vacuum properties}

Once we have fixed the model parametrization, we can calculate the values of several meson properties for the scalar and pseudoscalar sector. 
Our numerical results are summarized in Table~\ref{tab:results}, together with the corresponding phenomenological estimates. 
The quantities marked with an asterisk are those that have been chosen as inputs. 

In general, it is seen that meson masses, mixing angles and weak decay constants predicted by the model are in a reasonable agreement with phenomenological expectations.
\begin{table}[h]
\begin{center}
\begin{tabular*}{0.4\textwidth}{@{\extracolsep{\fill}} c c c }
\hline
\hline
 & Model & Empirical \\
\hline 
$\bar\sigma_u$ [MeV] & $400$ & ... \\
$\bar\sigma_s$ [MeV] & $630$ & ... \\
$\bar\zeta / \kappa$ & $-0.332$ & ... \\
$-\langle \overline{u} u \rangle ^{1/3}$ [MeV] & $325$ & ... \\
$-\langle \overline{s} s \rangle ^{1/3}$ [MeV] & $358$ & ... \\
\hline
$m_{\pi}$ [MeV] &  $139$ $^\ast$  & $139$ \\
$m_{\sigma}$ [MeV] & $518$ & $400 - 550$ \\
$m_{K}$ [MeV] & $495$ $^\ast$ & $495$ \\
$m_{K_0^\ast}$ [MeV] & $1159$ & $1425$ \\
$m_{\eta}$ [MeV] & $511$ & $547$ \\
$m_{{\rm a}_0}$ [MeV] & $968$ & $980$ \\
$m_{\eta^{\prime}}$ [MeV] & $958$ $^\ast$ & $958$ \\
$m_{{\rm f}_0}$ [MeV] & $1280$ & $990$ \\
\hline
$f_{\pi}$ [MeV] & $92.4$ $^\ast$ & $92.4$ \\
$f_K/f_{\pi}$ & $1.18$ & $1.22$ \\
$f_{\eta}^0/f_{\pi}$ & $0.27$ & $(0.11$ - $0.51)$ \\
$f_{\eta}^8/f_{\pi}$ & $1.05$ & $(1.17$ - $1.22)$ \\
$f_{\eta^{\prime}}^0/f_{\pi}$ & $2.12$ & $(0.98$ - $1.16)$ \\
$f_{\eta^{\prime}}^8/f_{\pi}$ & $-0.63$ & $-(0.42$ - $0.46)$\ \\
\hline
$\theta_{0}$ &  $-7^\circ$ & $-(10^\circ$ - $12^\circ)$ \\
$\theta_{8}$ &  $-31^\circ$ & $-(25^\circ$ - $29^\circ)$ \\
\hline
\hline
\end{tabular*}
\caption{\small{Numerical results for various phenomenological quantities. Input values are marked with an asterisk.}}
\label{tab:results}
\end{center}
\end{table}

As in precedent analyses~\cite{Noguera:2008,Hell:2011ic,Carlomagno:2013ona}, we obtain relatively low values for $m_u$ and $m_s$, and a somewhat large value for the light quark condensate. 
On the other hand, we find that the quark mass ratio is $m_s/m_u\simeq 26$, which is phenomenologically adequate. 
Something similar happens with the product $-\langle \bar uu\rangle m_u$, which gives $8.17\times 10^{-5}$~GeV$^4$, in agreement with the scale-independent result obtained from the Gell-Mann-Oakes-Renner relation at the leading order in the chiral expansion, namely $-\langle \bar uu\rangle m_u = f_\pi^2 m_\pi^2/2 \simeq 8.25\times 10^{-5}$~GeV$^4$.

Notice that the set of parameters quoted in Table~\ref{tab:param} differs from the one used in Ref.~\cite{Carlomagno:2013ona}. 
As it was explained by the authors in Ref.~\cite{Villafane:2016ukb}, the numerical evaluation of loop integrals has to be treated with some care given the functional form of the lattice inspired form factors, since for instance the function $f_m(s)$ in Eq.~(\ref{fm2}) presents branch cuts in the complex plane for ${\rm Re}(s)<0$,  ${\rm Im}(s)=0$.  
The presence of these cuts generates new contributions to the loop integrals, and therefore the values of the presented free parameters are different from those in Ref.~\cite{Carlomagno:2013ona}.

\subsection*{Finite temperature phenomenology}

In previous works~\cite{Carlomagno:2013ona,Carlomagno:2015hea} we have analyzed the thermal behavior of thermodynamic quantities such as entropy, energy density and interaction measure in this kind of models. 
Here, we will describe the temperature dependence of meson masses, mixing angles and decay constants, which has not been previously addressed in SU(3) nonlocal models with WFR and/or lQCD-inspired form factors.

In addition, in Ref.~\cite{Carlomagno:2013ona} we have studied the mentioned thermal properties for Gaussian form factors, which guarantee a fast ultraviolet convergence of the loop integrals.
However, this kind of exponential momentum dependence provides unfavorable predictions in comparison with lQCD estimations and results coming from the previously introduced lQCD-inspired form factors.
This same improvement is also appreciable in the temperature dependence of meson masses, mixing angles and decay constants presented in this section (see~\cite{Contrera:2009hk} for an analysis with Gaussian form factors).

\begin{figure}[h]
\centering
\includegraphics[width=0.45\textwidth]{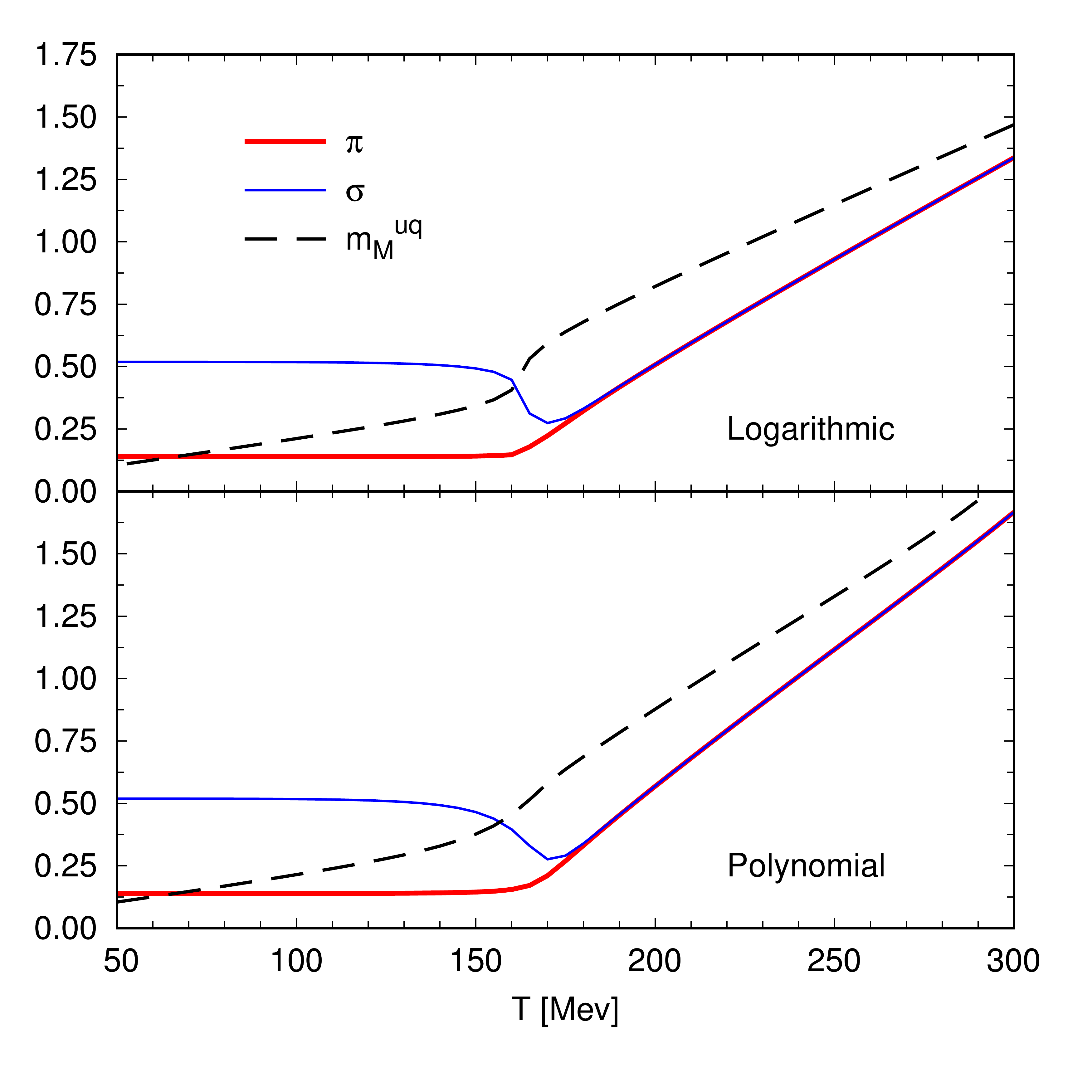}
\caption{\small{Pion (thick line) and sigma (thin line) mass as functions of $T$ for the logarithmic and polynomial potential in upper and lower panel, respectively. Dashed lines represent the thermal energy.}}
\label{fig:vs}
\end{figure}

In Fig.~\ref{fig:vs} we show the behavior of spatial masses of mesons $\sigma$ (thin line) and $\pi$ (thick line) as functions of the temperature, for the logarithmic (upper panel) and polynomial (lower panel) effective potentials given by Eqs.~(\ref{log}) and~(\ref{pol}), respectively.
Around the critical temperature, it is possible to distinguish a stronger steepness in the curves for the logarithmic potential.
In addition, the higher the temperature, the larger is the splitting between the  predictions for both PL potentials.
At high temperature the masses are dominated by the thermal energy (dashed lines), hence the behavior should approach that of an uncorrelated pair of massless quarks,  $m_M^{\rm uq} = 2 \pi T$~\cite{Eletsky:1988an}. 
However, for $T \sim 300$~MeV the trace of the Polyakov loop has not yet reached its asymptotic value $\Phi=1$, and there is still a non negligible contribution to the quark effective mass provided by the background color field $\phi$. 
Indeed, the thermal energy behaves as $m_M^{\rm uq} = 2(\pi T - \phi)$.

\begin{figure}[h]
\centering
\includegraphics[width=0.45\textwidth]{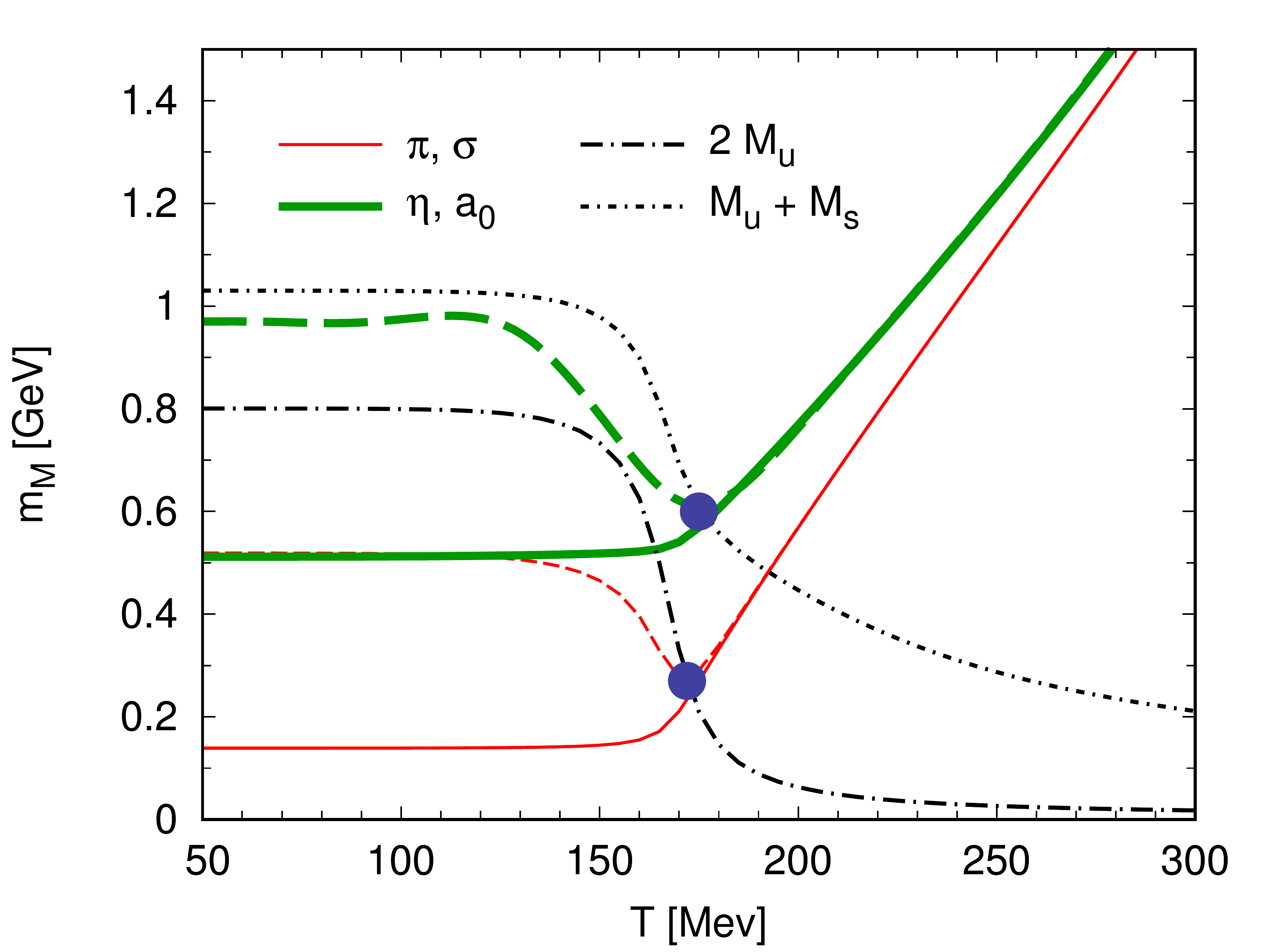}
\caption{\small{Scalar (dashed line) and pseudoscalar (solid line) meson masses as function of $T$ for the polynomial potential. Effective quark masses are plotted in dashed-dotted lines. The Mott temperature is indicated by the large dot.}}
\label{fig:masas_mott}
\end{figure}

\begin{figure}[h]
\centering
\includegraphics[width=0.45\textwidth]{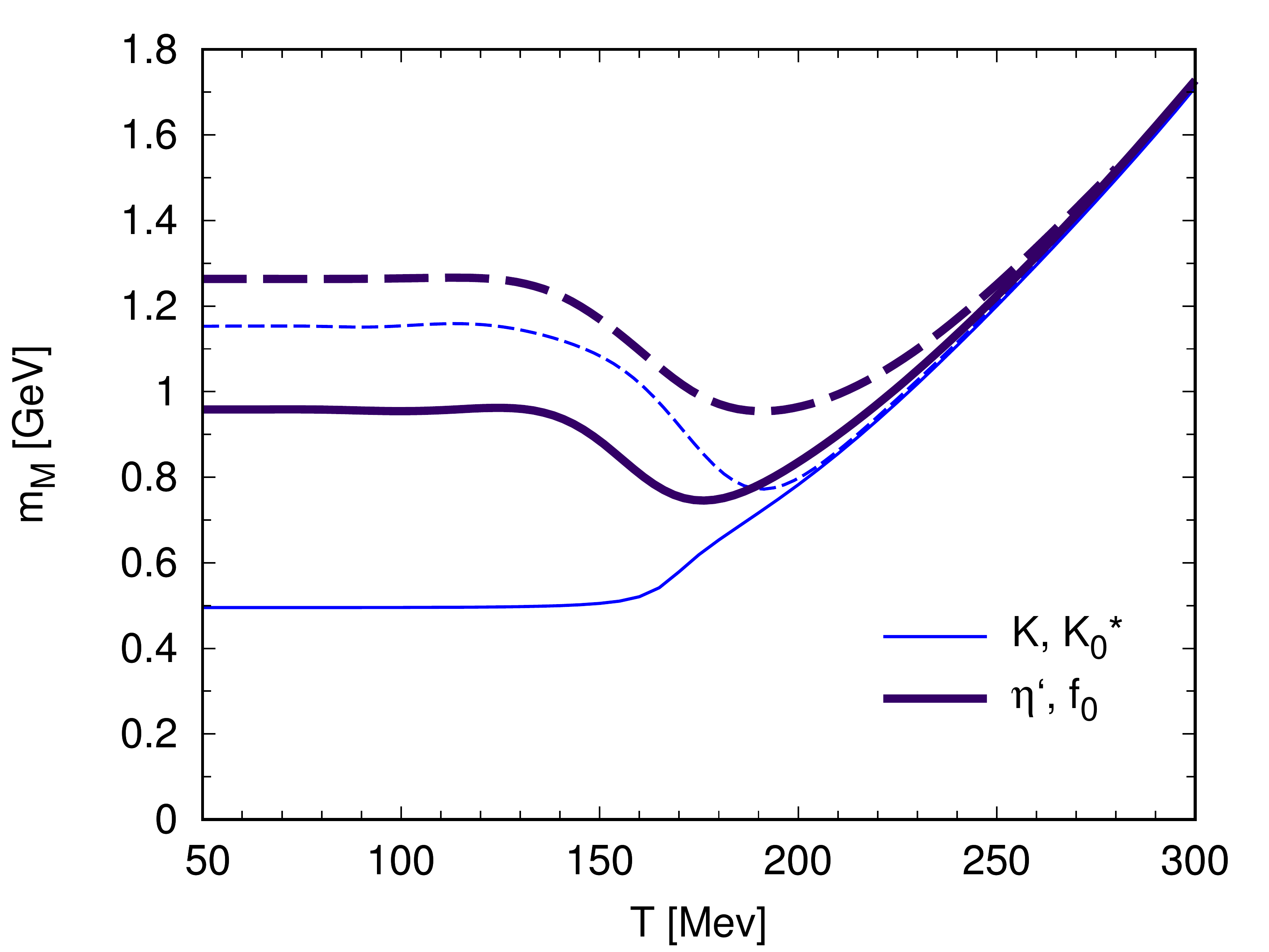}
\caption{\small{Scalar (dashed line) and pseudoscalar (solid line) meson masses as function of the temperature for the polynomial PL potential.}}
\label{fig:masas}
\end{figure}

\begin{figure}[h]
\centering
\includegraphics[width=0.45\textwidth]{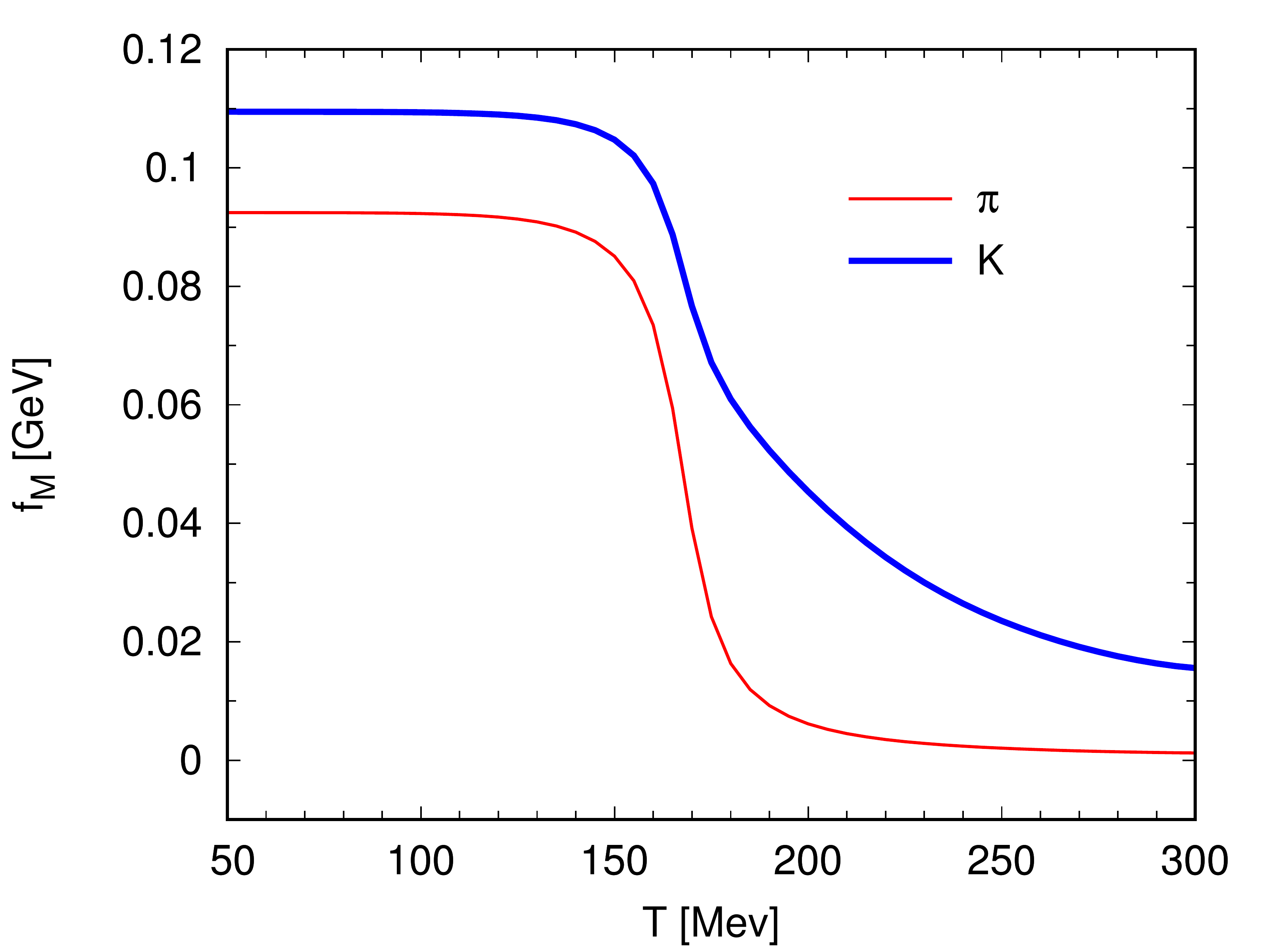}
\caption{\small{Pion (thin line) and kaon (thick line) decay constants as function of the temperature for the polynomial PL potential.}}
\label{fig:efes}
\end{figure}

It is known that the polynomial potential has a smoother behavior and reaches faster the deconfinement asymptotic value, in agreement with our results.
Nevertheless, the qualitative thermal evolution for meson masses, decay constants and mixing angles is similar for both potentials. 
Here we show the results for the polynomial potential, since it provides the best agreement with lQCD results for the chiral restoration transition~\cite{Carlomagno:2013ona}.
In Figs.~\ref{fig:masas_mott} and~\ref{fig:masas} we plot meson and constituent quark  masses as functions of the temperature, whereas in Fig.~\ref{fig:efes} we show the behavior of $\pi$ and $K$ weak decay constants.

Effective theories that do not include an explicit mechanism of confinement, like PNJL models, usually present a threshold above which constituent quarks can be simultaneously on shell.
This leads to an imaginary part of the effective mass that can be interpreted as the width of a decay of the meson into two massive quarks.
That threshold, which depends on the model parametrization, is typically of the order of $1$~GeV. 
Therefore, the description of bound states lying above this value requires some regularization. 
In the cases where the functions $G_{M}(-p^2,0)$ have no zeros for real values of $p$, we have defined the meson mass through the minimum of $\vert G_{M}(-p^2,0) \vert$.

It can be seen from Fig.~\ref{fig:masas_mott} and~\ref{fig:masas} that pseudoscalar meson masses (solid lines) remain approximately constant up to the critical temperature $T_c$, while scalar meson masses (dashed lines) start dropping somewhat below $T_c$.
Right above $T_c$, masses of chiral partners become degenerate.
Then, at higher temperatures, they are dominated by the thermal energy. 
In the case of the $\eta'$ meson and its chiral partner f$_0$, and similarly for $K$ and $K_0^*$, the degeneracy is achieved at larger temperatures than in the case of the other mesons (see Fig.~\ref{fig:masas}).
This a consequence of the strange quark content, which becomes larger compared to the content of other flavors as the temperature increases. 

In Fig.~\ref{fig:masas_mott}, besides the meson masses we plot in dashed-dotted lines our results for $2 M_u$ and $M_u + M_s$.
Up to certain temperature $T_m$ (denoted in the figure with a large dot) the mesons have a lower mass than the masses of their constituents. 
When $T>T_m$, meson masses are no longer a discrete solution of Eq.~(\ref{Ges}), which implies a passage from the discrete to the continuum, known as Mott transition~\cite{Blaschke:1984yj,Hufner:1996pq}. 
From  Fig.~\ref{fig:masas_mott} one can see that both Mott temperatures are $T_m \sim 170$~MeV. 
Above this temperature the meson should not be described by a bound state, but as a correlated state formed by a quark and an antiquark, which will deconfine when the temperature increases sufficiently.

It can be analytically proved and numerically checked that the branch cuts in the complex plane, coming from the momentum dependence of the nonlocal form factors, vanish when $T>T_m$. 
In this region, meson masses are lower than $m_M^{\rm uq}$ and therefore the necessary conditions (${\rm Re}(s)<0$ and ${\rm Im}(s)=0$) do not hold.
In other words, contributions from the cuts to the loop integrals are nonzero only when meson masses are discrete solutions of Eq.~(\ref{Ges}) and larger than $m_M^{\rm uq}$.

This section concludes with the analysis of the thermal behavior of pseudoscalar meson decay constants, quoted in Fig.~\ref{fig:efes}, and mixing angles $\theta_0$ and $\theta_8$.
For the former we see that pseudoscalar mesons with larger content of strangeness present a decay constant with a more moderate decrease after the transition.
For the mixing angles it is seen (as in Ref.~\cite{Contrera:2009hk}) that above $T_c$, $\theta_0$ and $\theta_8$ tend to a common value, the so-called ``ideal'' mixing angle $\theta_{\rm ideal}=\tan^{-1}\sqrt{2}\simeq 54.7^\circ$. 
This means that the $\eta$ meson becomes approximately non-strange, while $\eta'$ approaches to an $\bar ss$ pair.
The fact that the mixing angles go to the ``ideal'' value for large temperatures is related to the restoration of the  U(1)$_A$ symmetry.

\section{$T - \mu$ phase diagram} 
\label{pd}

In this section we discuss the features of phase transitions in the $T - \mu$ plane for the nonlocal chiral quark model introduced in Sect.~\ref{nlpnjl}. 
The phase diagram can be sketched by analyzing the numerical results obtained for the relevant order parameters.
In general, one can find regions in which the chiral symmetry is either broken or approximately restored through first order or crossover phase transitions, and phases in which the system remains either in confined or deconfined states. 

\begin{figure}[h]
\centering
\includegraphics[width=0.45\textwidth]{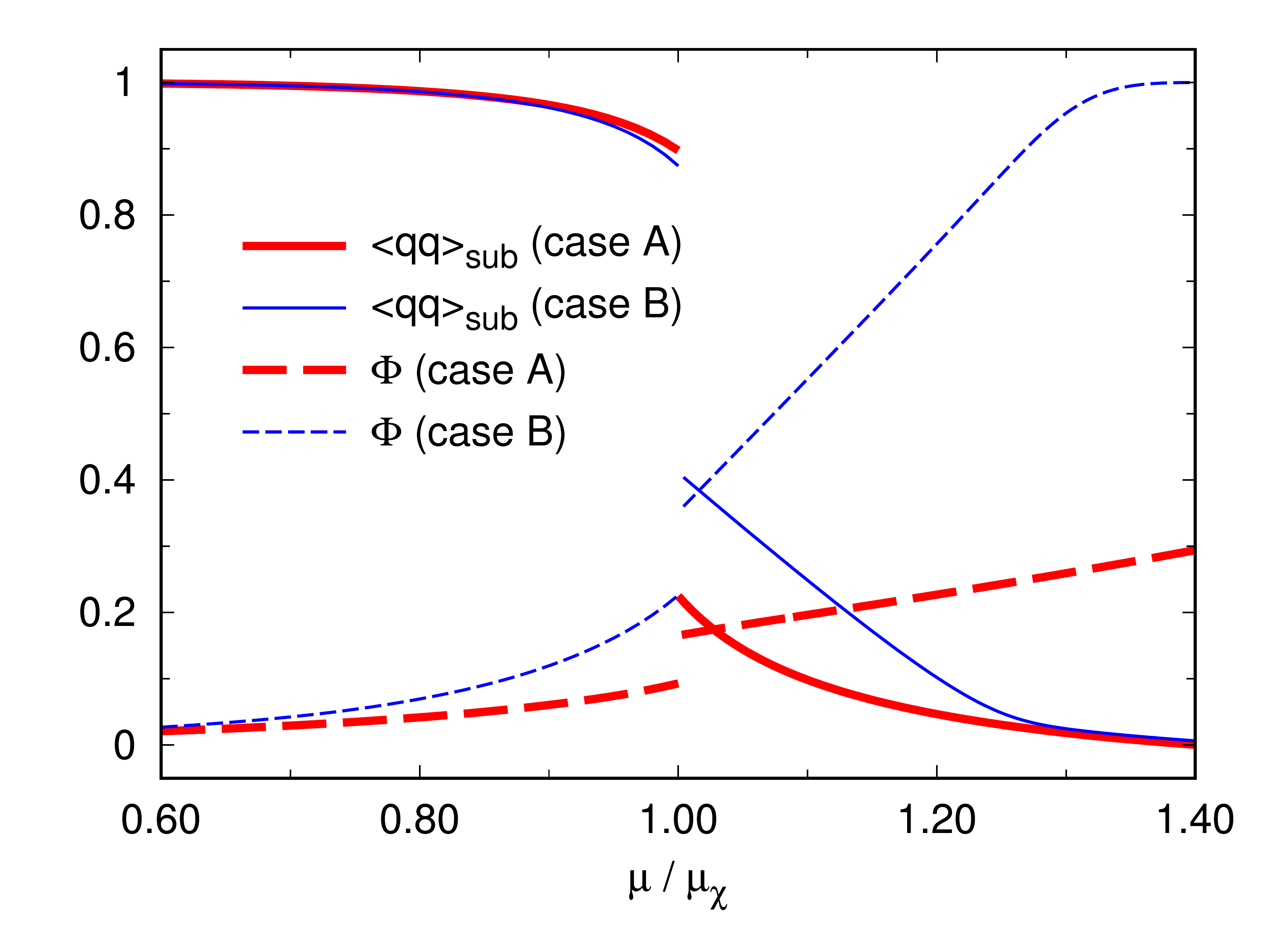}
\caption{\small{Trace of the Polyakov loop (dashed lines) and subtracted chiral condensate (solid lines) as a function of $\mu/\mu_{\chi}$, for $T=80$~MeV. Thick and thin lines correspond to constant and $\mu$-dependent $T_0$, respectively.}}
\label{fig:Tcero}
\end{figure}

To study the QCD phase diagram, we improve the analysis carried out in Refs.~\cite{Pagura:2011rt,Carlomagno:2013ona} for the dependence of critical temperatures and chemical potentials with the parameter $T_0$, by comparing two complementary scenarios. 
In the first one (say, situation \textbf{A}), we consider a constant value, $T_0=200$~MeV; in the second (situation \textbf{B}), we assume that $T_0$ depends on the chemical potential accoding to Eq.~(\ref{T0ofmu}).
The comparison between these two cases is illustrated in Fig.~\ref{fig:Tcero}, where we plot the subtracted chiral condensate (solid line) and the trace of the Polyakov loop (dashed line) as functions of the reduced chemical potential $\mu/\mu_\chi$, for a representative value of the temperature, namely $T=80$~MeV.
The results correspond to the polynomial PL potential.

For relatively high temperatures, chiral restoration takes place as a smooth crossover, whereas at low temperatures the order parameter has a discontinuity at a given critical chemical potential $\mu_\chi$ signaling a first order phase transition. 
This gap in the quark condensate induces also a jump in the trace of the PL.
The value of $\Phi$ at both sides of the discontinuity indicate if the system remains confined or not.
As it was explained in Sect.~\ref{nlpnjl}, values close to zero or to one correspond to confinement or deconfinement, respectively.
From Fig.~\ref{fig:Tcero} we can see that, for the chosen temperature $T=80$~MeV, at the critical chemical potential $\mu_{\chi}$ ($\mu/\mu_{\chi}=1$) the value of $\Phi$ for situation \textbf{B} (thin dashed line) is at least twice larger than for situation \textbf{A} (thick dashed line).
Moreover, in situation \textbf{B} the value can be assumed to be high enough to consider that quarks are no longer confined into hadrons, while in situation \textbf{A}, the system remains in a confined phase in which chiral symmetry is approximately restored.
This difference between both situations holds for all relevant values of the temperature.
Therefore, for a PL potential with $T_0$ given by Eq.~(\ref{T0ofmu}), the chiral restoration and the confinement-deconfinement transitions take place always simultaneously, in agreement with the analysis made in Ref.~\cite{Schaefer:2007pw} within a Polyakov-quark-meson model.

In heavy ion collisions it is believed that before the occurrence of the kinetic freeze out a mixed phase of quarks and hadrons could exist~\cite{Kumar:2013cqa}.
As discussed above, a $\mu$-dependent $T_0$ leads to a QCD phase diagram without such a mixed phase.
Therefore, we concentrate mainly on the case of a constant $T_0$, where for large densities and for a certain temperature range, where the chiral symmetry is restored, the trace of the Polyakov loop still indicates confinement.

As stated, for the deconfinement and chiral symmetry restoration transitions we take as order parameters the traced Polyakov loop $\Phi$ and the subtracted chiral condensate $\langle \bar q q\rangle_{\rm sub}$, respectively.
The associated susceptibilities $\chi_\Phi$ and $\chi_q$ are given by the derivatives
\begin{equation}
\chi_\Phi = \frac{d \Phi} {d T}  \quad{\rm and}\quad \chi_q = \frac{\partial^2\Omega^{\rm MF}}{\partial m_q^2} \ . \nonumber 
\end{equation} 

The associated critical temperatures $T_{\chi}$ and $T_\Phi$ are defined by the position of the peaks in the chiral susceptibilities in the region where the transition occurs as a smooth crossover. 

When the chiral restoration occurs as a first order phase transition, the PL susceptibility present a divergent behavior at the chiral critical temperature even when the order parameter $\Phi$ remains close to zero.
Therefore we need another definition for the deconfinement critical temperatures in this region of the phase diagram. 
Here, we employ the same prescription as in Ref~\cite{Contrera:2010kz}, namely, we define the critical temperature requiring that $\Phi$ takes a given value.
We choose a range between $0.3$ and $0.5$, which could be taken as large enough to denote deconfinement.

At zero temperature, the chiral restoration occurs through a first order phase transition at a critical chemical potential $\mu_\chi \sim 290$~MeV. 
If we move, in the $T-\mu$ plane, along the first order phase transition curve, the critical temperature rises from zero up to a critical endpoint (CEP) temperature $T_{CEP}$, while the critical chemical potential decreases from $\mu_{\chi}$ to $\mu_{CEP}$. 
Beyond this point, the chiral restoration phase transition proceeds as a smooth crossover. 
The numerical results for the CEP coordinates, critical temperatures and densities are summarized in Table~\ref{tab:cep}.
The positions in the $T,\mu$ plane of these critical points are similar to those obtained in Refs.~\cite{Contrera:2012wj,Carlomagno:2015hea} for nlPNJL models with two dynamical quark flavors.
In addition, to the best of our knowledge, this work is the first investigation within SU(3) nonlocal effective theories on the study of the QCD phase diagram.
\begin{table}[h]
\begin{center}
\begin{tabular*}{0.3\textwidth}{@{\extracolsep{\fill}} ccc }
\hline 
\hline 
 & Logarithmic & Polynomial \\
\hline 
$T_{\rm CEP}$   & 130 & 112 \\
$\mu_{\rm CEP}$ & 214 & 234 \\
\hline 
$T_c\ (\mu=0)$  & 163 & 169 \\
$T_{\chi}\ (\mu=100)$  & 158 & 161 \\
$T_{\chi}\ (\mu=250)$  & 108 & 98 \\
$T_{\chi}\ (\mu=280)$  & 71 & 64 \\
$\mu_{\chi}\ (T=0)$  & \multicolumn{2}{c}{293} \\
\hline 
\hline 
\end{tabular*}
\caption{\small{Critical temperatures and densities and CEP coordinates for both PL effective potentials.}}
\label{tab:cep}
\end{center}
\end{table}

On the other hand, at zero chemical potential, when the temperature increases, the system undergoes both chiral restoration and deconfinement transitions at a critical temperature of $T_c \sim 165$~MeV, which proceed as smooth crossovers, in agreement with lQCD. 
Moreover, in Ref~\cite{Bazavov:2016uvm} it is shown that the deconfinement temperature, defined by the peak of the entropy of a static quark (which is related to the Polyakov loop) coincides, within errors, with the chiral restoration temperature.

\begin{figure}[h]
\begin{center}
\includegraphics[width=0.45\textwidth]{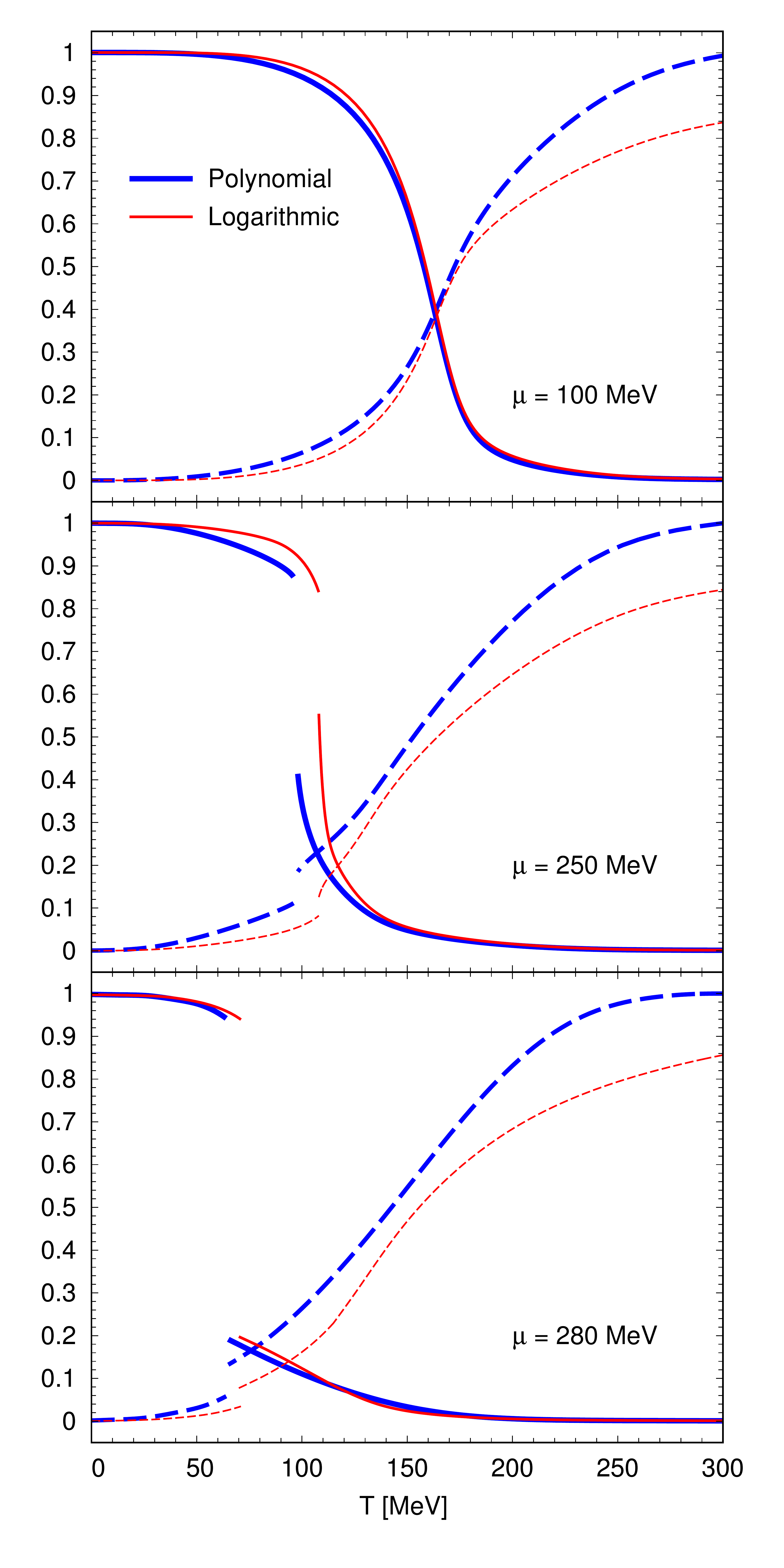}
\end{center}
\caption{\small{Subtracted chiral condensate (solid line) and traced Polyakov loop (dashed line) as functions of the temperature for the logarithmic (polynomial) PL potential in thin (thick) lines.}}
\label{fig:mu_cte}
\end{figure}

However, for chemical potentials larger than $\mu_{\rm CEP}$ these transitions begin to separate.
This can be seen in Fig.~\ref{fig:mu_cte}, where we quote for some given values of $\mu$ the order parameters for the deconfinement transition and the chiral symmetry restoration as functions of the temperature, plotted in dashed and solid lines respectively, for the logarithmic (thin line) and polynomial (thick line) effective potentials. 
In the upper panel, which corresponds to $\mu=100$~MeV, it is seen that the chiral and deconfinement transitions proceed as smooth crossovers occurring at the same critical temperature.
When the chemical potential becomes larger than $\mu_{\rm CEP}$ (see Table~\ref{tab:cep}), the order parameter for the chiral symmetry restoration has a discontinuity signaling a first order phase transition. 
These gap in the quark condensate induces also a jump in the trace of the PL (see central and lower panels in Fig.~\ref{fig:mu_cte}, where $\mu=250$~MeV and $\mu=280$~MeV, respectively).
The relatively low values of $\Phi$ at the discontinuity indicate that after the transition the system remains confined but in a chiral symmetry restored state. 
The deconfinement occurs at larger temperatures when the order parameter becomes closer to one. 
The phase in which quarks remain confined (signaled by $\Phi\lesssim 0.3$) even though chiral symmetry has been restored is usually referred to as a quarkyonic phase~\cite{McLerran:2007qj,McLerran:2008ua,Abuki:2008nm}.

\begin{figure*}[t]
\begin{center}
\includegraphics[width=0.95\textwidth]{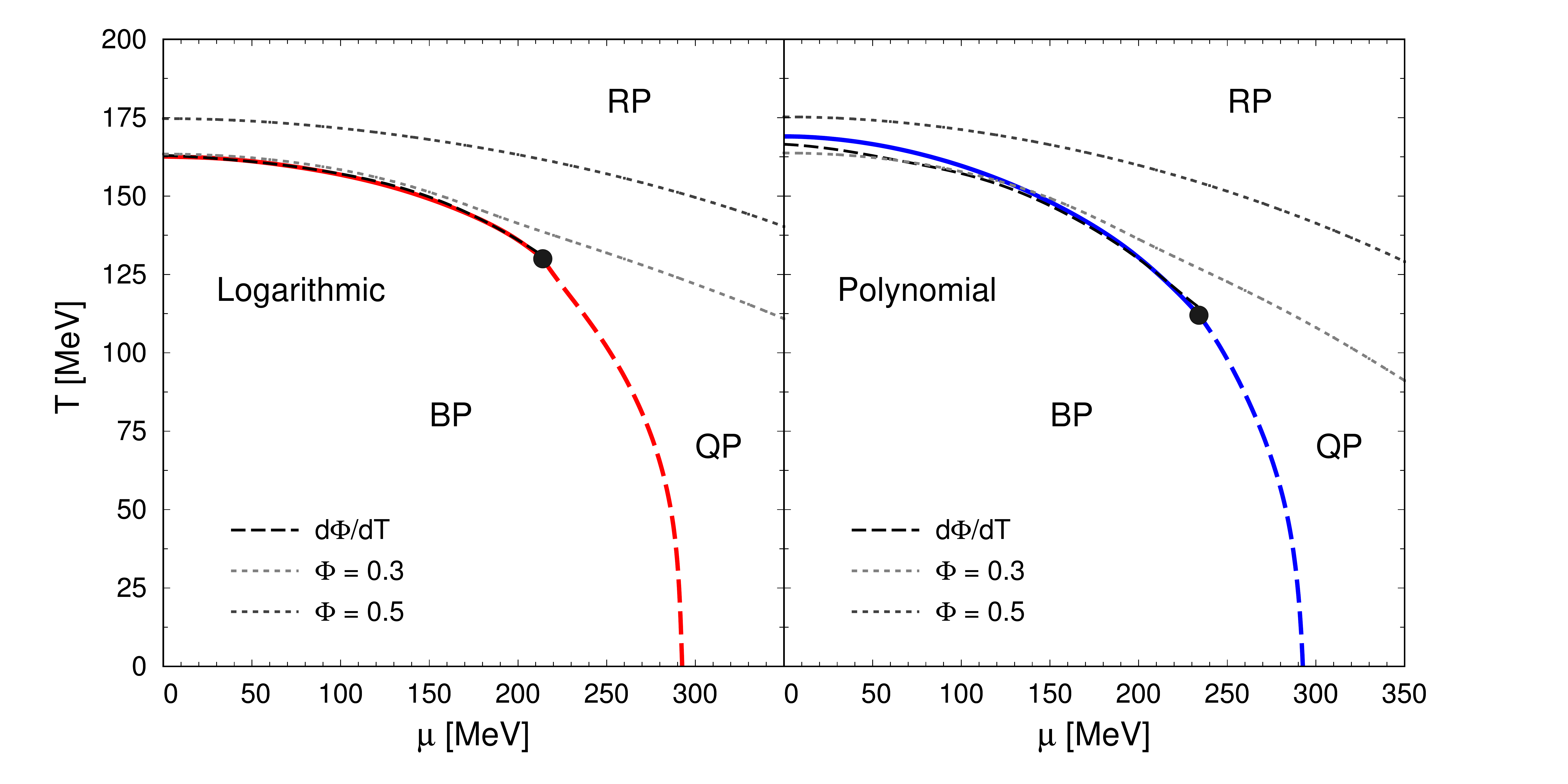}
\end{center}
\caption{\small{Phase diagrams for logarithmic (left panel) and polynomial (right panel) PL potentials. The CEP is denoted by the dot. Dashed and solid lines indicate crossover and first order chiral transitions, respectively. Dotted lines correspond to deconfinement transitions.}}
\label{fig:pd_SU3}
\end{figure*}

We quote in Fig.~\ref{fig:pd_SU3} the phase diagrams for the $SU(3)$ nonlocal PNJL model described in Sect.~\ref{nlpnjl} considering both logarithmic and polynomial PL potentials, on left and right panels respectively.
In solid (dashed) lines we plot first order (crossover) phase transitions for the chiral symmetry restoration, while the deconfinement transition lines defined by $\Phi=0.3$ and $\Phi=0.5$ are plotted in dotted lines.
The dot denotes the position of the critical endpoint. 

At a given chemical potential lower than $\mu_{\chi}$, when the temperature increases one finds a transition from a hadronic phase with broken chiral symmetry (BP), to a quarkyonic phase (QP) where the chiral symmetry is restored but the quarks are still confined into hadrons. 
If the temperature continues raising, the deconfinement transition takes place and one reaches a partonic phase in which the quarks are deconfined and the chiral symmetry is restored (RP).

\section{Summary and conclusions}
\label{summary}

Along this work we have studied light scalar and pseudoscalar meson properties and the characteristics of deconfinement and chiral restoration transitions in the context of a three-flavor nonlocal chiral model. 
Gauge interactions have been effectively introduced through a coupling between quarks and a constant background color gauge field, the Polyakov field, whereas gluons self-interactions have been implemented through logarithmic and polynomial effective Polyakov loop potentials.
The analysis done in this article should be endorsed as extensions of previous works, Refs.~\cite{Contrera:2009hk,Carlomagno:2013ona}. 

Within this framework we have obtained a parametrization that reproduces lattice QCD results for the momentum dependence of the effective quark mass and WFR, and at the same time leads to an acceptable phenomenological pattern for particle masses and decay constants in both scalar and pseudoscalar meson sectors.
In our calculations we have included the contributions from branch cuts in the momentum complex plane that arise from the lattice inspired nonlocal form factors.

As a second step, we have analyzed the temperature dependence of several meson properties, like meson masses, decay constants and mixing angles.
As expected, it is found that meson masses get increased beyond the chiral critical temperature, becoming degenerated with their chiral partners. 
The temperatures at which this happens depend on the strange quark content of the corresponding mesons.

Meson masses and weak decay constants remain approximately constant up to the critical chiral temperature.
In addition, light hadrons with strange degrees of freedom present a decay constant with a less steep decrease. 
Regarding the properties of the mixing angles, they tend to converge to the so-called ``ideal'' mixing, which indicates that the U(1)$_A$ anomaly tends to vanish as the temperature increases.

Finally, we study the characteristics of deconfinement and chiral restoration transitions at finite temperature and chemical potential.
As expected, at zero $\mu$, the model shows a crossover phase transition, corresponding to the restoration of the SU(2) chiral symmetry. 
The transition temperature is found to be $T_c \sim 165$~MeV, in very good agreement with lattice results. 
In addition, one finds a deconfinement phase transition, which occurs at the same critical temperature.
On the other hand, at zero temperature chiral restoration takes place via a first order transition at a critical density $\mu_\chi \sim 290$~MeV, in agreement with estimations coming from compact objects.

For chemical potentials larger than $\mu_{\rm CEP}$, the critical temperatures for the restoration of the chiral symmetry and deconfinement transition begin to separate. The region between them denotes a phase where the chiral symmetry is restored but quarks remains confined, known as quarkyonic phase.
This splitting is strongly dependent of the parameter $T_0$ entering in the PL potential.
If we consider for this parameter an explicit dependence with $\mu$, both transitions are always simultaneous, and therefore there is no such mixed phase, in contradiction with some results from heavy-ion collisions.

\section*{Acknowledgements}
I would like to thank D.~G{\'o}mez Dumm for a detailed reading of the manuscript and N.~N.~Scoccola for helpful discussions.
This work has been partially supported by CONICET under grants P-UE 2016 ``B{\'u}squeda de Nueva F{\'i}sica'' and  PIP12-449, and by the National University of La Plata, Project No.~X718.

\pagebreak

\end{document}